\documentclass[a4paper,11pt]{article}

\usepackage{amsmath}
\usepackage{amssymb}
\usepackage[numbers, square, comma, sort&compress]{natbib}
\usepackage{hyperref}
\usepackage{graphicx}
\usepackage{epstopdf}
\usepackage{bm}
\usepackage{epsfig}
\usepackage{graphics}
\usepackage{xspace}
\usepackage{amsfonts}
\usepackage{verbatim}
\usepackage{geometry}
\usepackage{color}
\usepackage{array} 
\usepackage{comment}
\usepackage{subfiles}
\usepackage{multirow}
\usepackage{soul}
\usepackage[none]{hyphenat}
\graphicspath{{figures/}}
\usepackage{tikz-feynman}
\tikzfeynmanset{compat=1.0.0}

\allowdisplaybreaks

\begin{document}

\begin{titlepage}
\begin{center}
\Large\bf{Improved constraints on Higgs boson self-couplings with quartic and cubic power dependencies of the cross section}
\end{center}
\vspace{1cm}

\begin{center}
{\bf Hai Tao Li$^a$, Zong-Guo Si$^a$, Jian Wang$^{a,b}\footnote{j.wang@sdu.edu.cn}$,  Xiao Zhang$^a$, Dan Zhao$^a$}
\vspace{1cm}

\textit{$^a$School of Physics, Shandong University, Jinan, Shandong 250100, China} \\ 
\vspace{1mm}
\textit{$^b$Center for High Energy Physics, Peking University, Beijing 100871, China} \\

\end{center}

\vspace{10mm}

\begin{abstract}
\begin{sloppypar}

Precise determination of the Higgs boson self-couplings is essential for understanding the mechanism
underlying electroweak symmetry breaking. However, owing to the limited number of Higgs boson pair events at the
LHC, only loose constraints have been established to date. Current constraints are based on the assumption that the
cross section is a quadratic function of the trilinear Higgs self-coupling within the $\kappa$ framework. 
Incorporating higher-order quantum corrections from virtual Higgs bosons would significantly alter this functional form, introducing
new quartic and cubic power dependencies on the trilinear Higgs self-coupling. To derive this new functional form,
we propose a specialized renormalization procedure that tracks all Higgs self-couplings at each calculation step. 
Additionally, we introduce renormalization constants for coupling modifiers within the $\kappa$ framework to ensure the cancellation of all ultraviolet divergences.
With new functional forms of the cross sections in both the gluon-gluon fusion and vector boson fusion channels, the upper limit of $\kappa_{\lambda_{\rm 3H}}=\lambda_{\rm 3H}/\lambda_{\rm 3H}^{\rm SM}$ set by the ATLAS (CMS) collaboration is reduced from 6.6 (6.49) to  5.5 (5.39).
However, extracting a meaningful constraint on the quartic Higgs self-coupling $\lambda_{\rm 4H}$ from Higgs boson pair production data remains challenging.
We also present the invariant mass distributions of the Higgs boson pair at different values of $\kappa_{\lambda}$, which could assist in setting optimal cuts for experimental analysis.

\end{sloppypar}

\vspace{1cm}


\end{abstract}

\end{titlepage}

\begin{sloppypar}

\section{Introduction}

Following the discovery of the Higgs boson at the Large Hadron Collider (LHC)~\cite{ATLAS:2012yve, CMS:2012qbp}, precise measurements of its properties, including mass, spin, and couplings to gauge bosons and fermions, have become critically important \cite{CMS:2020xrn, ATLAS:2022zjg, CMS:2022ley, ATLAS:2015zhl, CMS:2014nkk, CMS:2022uhn, ATLAS:2024vxc}.  
These measurements have thus far been consistent with the expectations of the Standard Model (SM)~\cite{ATLAS:2022vkf, CMS:2022dwd}.  However, the trilinear and quartic Higgs self-couplings, denoted as $\lambda_{\rm 3H}$ and $\lambda_{\rm 4H}$, respectively,
which represent a fundamental aspect of the SM that connects the Higgs mechanism and the stability of our Universe \cite{Degrassi:2012ry}, are still subject to large uncertainties.

Significant efforts have been dedicated to improving the measurement of the Higgs self-coupling. The most direct approach involves measuring the cross section of Higgs boson pair production, predominantly through gluon-gluon fusion (ggF). 
At the leading order (LO), the process occurs via a top-quark loop.
In the large top-quark mass ($m_t$) limit, the cross section of ggF Higgs boson pair production is known up to next-to-next-to-next-to-leading-order (N$^{3}$LO) QCD corrections ~\cite{Dawson:1998py,deFlorian:2013jea,deFlorian:2016uhr,Chen:2019lzz,Chen:2019fhs}, with the calculation of the soft gluon resummation effect also being studied \cite{Shao:2013bz,deFlorian:2015moa,Ajjath:2022kpv}.  
When considering the full $m_t$ dependence,
only next-to-leading-order (NLO) QCD corrections are  available~\cite{Borowka:2016ehy, Borowka:2016ypz, Baglio:2018lrj, Baglio:2020ini}, while estimates of the finite $m_t$ effects at next-to-next-to-leading-order (NNLO) have been conducted \cite{Grazzini:2018bsd,Czakon:2020vql,Mazzitelli:2022scc,Davies:2023obx,Davies:2024znp}.
A comprehensive simulation of events necessitates a fully differential calculation of the Higgs boson pair production and decay to the $b\bar{b}\gamma\gamma$ final state at QCD NLO~\cite{Li:2024ujf}.
Furthermore, investigations into the NLO electroweak corrections have been explored~\cite{Borowka:2018pxx, Muhlleitner:2022ijf, Davies:2022ram, Davies:2023npk, Bi:2023bnq,Heinrich:2024dnz,Davies:2024cxd}.
The subdominant channel---vector boson fusion (VBF)---has also been computed up to N$^3$LO in QCD \cite{Frederix:2014hta,Ling:2014sne,Dreyer:2018qbw,Dreyer:2018rfu,Dreyer:2020xaj}. 

The current constraints on the trilinear Higgs self-coupling extracted from the Run 2 dataset of Higgs boson pair production at the LHC by the ATLAS and CMS experiments are
$-0.6<\kappa_{\lambda_{\rm 3H}}<6.6$~\cite{ATLAS:2022jtk} and $-1.24 < \kappa_{\lambda_{\rm 3H}}< 6.49$~\cite{CMS:2022dwd}, respectively, within the $\kappa$ framework \cite{LHCHiggsCrossSectionWorkingGroup:2013rie},
where $\kappa_{\lambda_{\rm 3H}}=\lambda_{\rm 3H}/\lambda_{\rm 3H}^{\rm SM}$ with $\lambda_{\rm 3H}^{\rm SM}$ being the SM value of the trilinear Higgs self-coupling. 
On the other hand, the process of single Higgs boson production and decay depends on the Higgs self-coupling only via higher-order electro-weak (EW) corrections and can also impose certain constraint~\cite{McCullough:2013rea,Gorbahn:2016uoy,Degrassi:2016wml,Bizon:2016wgr,DiVita:2017eyz,Gao:2023bll}.
A measurement using the differential fiducial cross section in bins of the Higgs boson transverse momentum sets a constraint of $-5.4<\kappa_{\lambda_{\rm 3H}}<14.9$~\cite{CMS:2023gjz}.
Combined analyses of single- and double-Higgs production result in the constraint, $-0.4<\kappa_{\lambda_{\rm 3H}}<6.3$, at the $95\%$ confidence level, assuming that the new physics changes only the Higgs self-coupling~\cite{ATLAS:2022jtk}.

These constraints stem from considering the cross section of Higgs boson pair production as a function of the Higgs self-coupling.
Indeed, even with higher-order QCD corrections, the cross section is a quadratic function of the trilinear Higgs self-coupling $\lambda_{\rm 3H}$.
Nevertheless,
higher-order EW corrections introduce contributions from Feynman diagrams containing one or more triple Higgs or quadruple Higgs vertices, leading to a distinct functional dependence on the Higgs self-coupling. 
Specifically, new quartic and cubic power dependencies on the trilinear Higgs self-coupling emerge, which has a significant impact on the constraints, given that the current upper limit is so large.

In practical calculations, maintaining an explicit dependence on the Higgs self-coupling can be challenging, as it is typically treated as a derived parameter in the conventional calculation of EW corrections, particularly during the renormalization process \cite{Denner:1991kt, Bi:2023bnq}.
Even if one can perform the renormalization by taking the Higgs self-coupling as a primary parameter,
it is not clear how to implement the relation $m_H^2=2\lambda v^2$ and to rescale the Higgs self-coupling.
Different choices lead to different expressions for the cross sections.
Therefore, the calculation of the EW corrections in the SM can not be extended to the case with general $\lambda_{\rm 3H}$ and $\lambda_{\rm 4H}$,
and the cross section with higher power (beyond quadratic) dependence on the Higgs self-couplings in the general $\kappa$ framework is still lacking. 

To address this challenge, we propose a renormalization procedure that explicitly retains the Higgs self-couplings at each step and introduce  renormalization of the coupling modifier.
By combining it with the analytical and numerical calculations of the complex one-loop and two-loop amplitudes, respectively, we derive the cross sections of the Higgs boson pair production in both the ggF and VBF channels as functions of the Higgs self-couplings.
Our findings indicate that incorporating higher power dependencies of the cross section on the Higgs self-couplings can reduce the upper limit of $\kappa_{\lambda_{\rm 3H}}$ by approximately 20\%.

\section{Renormalization in the $\kappa$ framework} 
\label{method}

The LO contribution to the ggF Higgs boson pair production $g(p_1)g(p_2)\to H(p_3)H(p_4)$ arises from the top-quark induced triangle and box Feynman diagrams, which are of order $\lambda_{\rm 3H}$ and $\lambda_{\rm 3H}^0$, respectively. 
Therefore, the LO cross section at the 13 TeV LHC can be expressed as
\begin{align}
       \sigma_{\rm ggF,LO}^{\kappa_{\lambda}}=(~4.72~\kappa_{\lambda_{\rm 3H}}^2-23.0~\kappa_{\lambda_{\rm 3H}}+35.0  ~)~~ {\rm fb}.
       \label{eq:lokappa}
\end{align}
Here, the subscript $\lambda_{\rm 3H}$ in $\kappa_{\lambda_{\rm 3H}}$ signifies the deviation of the trilinear Higgs self-coupling from its SM value.
Below we will also introduce $\kappa_{\lambda_{\rm 4H}}$ to denote the modification of the quartic Higgs self-coupling.
The SM cross section is recovered when $\kappa_{\lambda_{\rm 3H}}=1$ and $\kappa_{\lambda_{\rm 4H}}=1$.
This quadratic functional form persists even with the inclusion of higher-order QCD corrections \cite{DiMicco:2019ngk}, e.g.,
\begin{equation}
  \sigma_{\rm ggF, NNLO-FT}^{\kappa_{\lambda}}=(~10.8~\kappa_{\lambda_{\rm 3H}}^2-49.6~\kappa_{\lambda_{\rm 3H}}+ 70.0~)~~ {\rm fb}.
  \label{eq:nnlokappa}
\end{equation}
In this expression, the full one-loop real contributions are merged with other NNLO QCD corrections in the large $m_t$ limit.
It is worth noting that the effects of QCD corrections are substantial, with each term in $\kappa_{\lambda_{\rm 3H}}$ more than doubling compared to the LO expression.

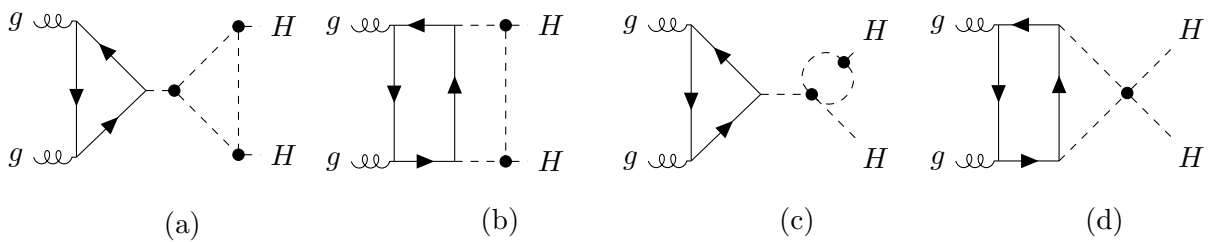
\begin{figure}[ht]
	\centering
	\begin{minipage}{0.23\linewidth}
		\centering
		\begin{tikzpicture}
		\begin{feynman}
		\vertex (i1) {\(g\)};
		\vertex[right=0.8 cm of i1] (a);
		\vertex[below right=1.3 cm of a] (b);
		\vertex[right=0.3 cm of b, dot] (d) {};
		\vertex[above right=1.2 cm of d, dot] (e) {};
		\vertex[below right=1.2 cm of d, dot] (f) {};
		\vertex[right=0.6 cm of e] (f1) {\(H\)};
		\vertex[right=0.6 cm of f] (f2) {\(H\)};
		
		\vertex[below=1.8 cm of i1] (i2) {\(g\)};
		\vertex[right=0.8 cm of i2] (c);
		
		\diagram* {
			(i1) -- [gluon] (a) -- [anti fermion] (b) -- [scalar] (d) -- [scalar] (e) -- [scalar] (f1),
			(i2) -- [gluon] (c) -- [fermion] (b),
			(c) -- [anti fermion] (a),
			(d) -- [scalar] (f) -- [scalar] (f2),
			(e) -- [scalar] (f)};
		\node at (2.2,-2.7) {(a)};
		\end{feynman}
		\end{tikzpicture}
	\end{minipage}\quad
	\begin{minipage}{0.23\linewidth}
		\centering
		\begin{tikzpicture}
		\begin{feynman}
		\vertex (i1) {\(g\)};
		\vertex[right=0.8 cm of i1] (a);
		\vertex[right=0.8 cm of a] (b);
		\vertex[right=0.6 cm of b, dot] (e) {};
		\vertex[right=0.6 cm of e] (f1) {\(H\)};
		
		\vertex[below=1.8 cm of i1] (i2) {\(g\)};
		\vertex[right=0.8 cm of i2] (c);
		\vertex[right=0.8 cm of c] (d);
		\vertex[right=0.6 cm of d, dot] (f) {};
		\vertex[right=0.6 cm of f] (f2) {\(H\)};
		
		\diagram* {
			(i1) -- [gluon] (a) -- [anti fermion] (b) -- [scalar] (e) -- [scalar] (f1),
			(i2) -- [gluon] (c) -- [fermion] (d) -- [scalar] (f) -- [scalar] (f2),
			(a) -- [fermion] (c),
			(b) -- [anti fermion] (d),
			(e) -- [scalar] (f)};
		\node at (2.2,-2.6) {(b)};
		\end{feynman}
		\end{tikzpicture}
	\end{minipage}\quad
 \begin{minipage}{0.23\linewidth}
		\centering
		\begin{tikzpicture}
		\begin{feynman}
		\vertex (i1) {\(g\)};
		\vertex[right=0.8 cm of i1] (a);
		\vertex[below right=1.3 cm of a] (b);
		\vertex[right=0.6 cm of b, dot] (d) {};
		\vertex[above right=0.6 cm of d, dot] (e) {};
		\vertex[above right=0.6 cm of e] (f1) {\(H\)};
		\vertex[below right=1.2 cm of d] (f2) {\(H\)};
		
		\vertex[below=1.8 cm of i1] (i2) {\(g\)};
		\vertex[right=0.8 cm of i2] (c);
		
		\diagram* {
			(i1) -- [gluon] (a) -- [anti fermion] (b) -- [scalar] (d)[dot] -- [scalar, half left] (e)[dot] -- [scalar, half left] (d),
			(i2) -- [gluon] (c) -- [fermion] (b),
			(c) -- [anti fermion] (a),
			(d) -- [scalar] (f2),
			(e) -- [scalar] (f1)};
		\node at (2.2,-2.6) {(c)};
		\end{feynman}
		\end{tikzpicture}
	\end{minipage}\quad
	\begin{minipage}{0.23\linewidth}
		\centering
		\begin{tikzpicture}
		\begin{feynman}
		\vertex (i1) {\(g\)};
		\vertex[right=0.8 cm of i1] (a);
		\vertex[right=0.8 cm of a] (b);
		\vertex[below right=1.2 cm of b, dot] (e) {};
		\vertex[above right=1.2 cm of e] (f1) {\(H\)};
		\vertex[below right=1.2 cm of e] (f2) {\(H\)};
		
		\vertex[below=1.8 cm of i1] (i2) {\(g\)};
		\vertex[right=0.8 cm of i2] (c);
		\vertex[right=0.8 cm of c] (d);
		
		\diagram* {
			(i1) -- [gluon] (a) -- [anti fermion] (b) -- [scalar] (e)[dot] -- [scalar] (f1),
			(i2) -- [gluon] (c) -- [fermion] (d) -- [scalar] (e) -- [scalar] (f2),
			(a) -- [fermion] (c),
			(b) -- [anti fermion] (d)};
		\node at (2.2,-2.6) {(d)};
		\end{feynman}
		\end{tikzpicture}
	\end{minipage}
	\caption{Typical two-loop Feynman diagrams of order $\lambda^3_{\rm 3H}~(a), \lambda^2_{\rm 3H}~(b), \lambda_{\rm 4H} \lambda_{\rm 3H} ~(c)$ and $\lambda_{\rm 4H}~(d) $, respectively.}
	\label{figNLO_HH}
\end{figure}

The EW corrections would change the above function form in two aspects.
Firstly, the coefficients of the quadratic, linear, and constant terms are altered by the corrections induced by virtual gauge bosons.
However, the overall impact is relatively minor, typically amounting to only a few percent, as reported in Ref. \cite{Bi:2023bnq}.
Given this negligible influence on the constraints related to the Higgs self-coupling, these corrections are deemed insignificant for the purposes of our study and are consequently omitted. 
Secondly, higher power or new dependence on the Higgs self-couplings arises due to the corrections induced by virtual Higgs bosons. 
One can see in Fig.~\ref{figNLO_HH} some typical two-loop Feynman diagrams, 
which give contributions of order $\lambda_{\rm 3H}^3$, $\lambda_{\rm 3H}^2$, $\lambda_{\rm 4H} \lambda_{\rm 3H}$,  $\lambda_{\rm 4H}$ to the amplitudes.
As a result, the cross section now contains new quartic and cubic powers of $\lambda_{\rm 3H}$ and starts to be sensitive to the quartic Higgs self-coupling $\lambda_{\rm 4H}$. 
It is the goal of our work to assess these corrections, denoted by $\delta \sigma_{\rm EW}^{\kappa_\lambda}$, and their impact on the constraints on the Higgs self-couplings.

Our calculation of the two-loop diagrams proceeds as follows.
We use \textsc{FeynArts} \cite{Hahn:2000kx} to generate the Feynman diagrams and corresponding amplitudes.
The amplitudes are written as a linear combination of two tensor structures with the coefficients called form factors \cite{Plehn:1996wb}.
After performing the Dirac algebra with \textsc{FeynCalc} \cite{Mertig:1990an,Shtabovenko:2016sxi,Shtabovenko:2020gxv}, we are left with scalar integrals for each form factor.
Rather than endeavoring to reduce all scalar integrals to master integrals and establishing differential equations for these master integrals, we opt to directly compute the scalar integrals for specific phase space points employing the numerical package \textsc{AMFlow} \cite{Liu:2017jxz,Liu:2022chg}. 
This decision is motivated by the intricate nature of constructing differential equations, particularly with general kinematic dependencies, which can be exceedingly time-consuming.
Even if the differential equation is obtained, the analytical solution seems not feasible with current technologies because all the propagators are massive. 
Numerical solutions of the differential equations often suffer from accuracy loss. 
By contrast, direct numerical calculation at each phase space point ensures accuracy. 
The primary challenge lies in covering the entire phase space efficiently.
Fortunately, the process of $gg\to HH$ is dominated by the S-wave scattering, making the amplitude insensitive to the scattering angle.
Moreover, its dependence on the scattering energy is also weak, except in very high-energy regions, as illustrated below. These characteristics allow us to generate a grid with a limited data set, which can be used to accurately calculate the amplitude at any point in the phase space.

Notice that the sum of all one-particle irreducible two-loop diagrams is finite.
But the sum of one-particle reducible two-loop diagrams contains ultraviolet divergences.
They will cancel after considering the contributions from the counter-terms in renormalization.

In the SM, the Lagrangian for the Higgs sector can be written as
\begin{equation}
	\mathcal{L}_{\rm H}=(D_{\mu}\phi_0)^{\dagger}(D^{\mu}\phi_0)+\mu_0^2(\phi_0^{\dagger}\phi_0)-\lambda_0(\phi_0^{\dagger}\phi_0)^2,
 \label{eq:lagr}
\end{equation}
where $\phi_0$ denotes the bare Higgs doublet and $D_{\mu}$ is the covariant derivative. 
The relations between the bare fields and couplings, and their renormalized counterparts, are given by $\phi_0  = Z^{1/2}_{\phi}\phi\,, 
    \mu^2_0 = Z_{\mu^2}\mu^2\,, $ and $
    \lambda_0 = Z_{\lambda}\lambda.$

The EW gauge symmetry is spontaneously broken once the Higgs field develops a non-vanishing vacuum expectation value $v$. Taking the unitary gauge, we write the Higgs field as
\begin{equation}
    \phi=\frac{1}{\sqrt{2}}
         \left(
          \begin{array}{c} 
          0 \\
	     H + Z_v v
          \end{array}
          \right),
\end{equation} 
where $Z_v$ is the renormalization constant for the vacuum expectation value. 
The  renormalized Lagrangian in the $\kappa$ framework after EW gauge symmetry breaking is given by
\begin{align}
      \mathcal{L}_{\rm H}^{\kappa}&=\frac{1}{2}Z_{\phi}(\partial_{\mu}H)^2-\left(-\frac{1}{2}Z_{\mu^2}Z_{\phi}Z_v^2 \mu^2 v^2+\frac{1}{4}Z_{\lambda}Z_{\phi}^2 Z_v^4 \lambda v^4\right)
      -(Z_{\lambda}Z_{\phi}^2 Z_v^3 \lambda v^3-Z_{\mu^2}Z_{\phi} Z_v \mu^2 v)H
      \nonumber \\
       &-\left(\frac{3}{2}Z_{\lambda}Z_{\phi}^2 Z_v^2 \lambda v^2-\frac{1}{2}Z_{\mu^2}Z_{\phi}\mu^2\right)H^2
      -Z_{\kappa_{\rm 3H} } Z_{\lambda}Z_{\phi}^2 Z_v \lambda_{\rm 3H} v H^3
      -\frac{1}{4}Z_{\kappa_{\rm 4H} } Z_{\lambda}Z_{\phi}^2\lambda_{\rm 4H} H^4+\cdots\,,
      \label{eq:Lsm}
\end{align}
where the ellipsis represents the terms involving EW gauge bosons.
Note that the letter 
 $\lambda$ is solely used for the Higgs self-coupling in the SM
 but $\lambda_{\rm 3H}\equiv \kappa_{\lambda_{\rm 3H}}\lambda $ and $\lambda_{\rm 4H}\equiv \kappa_{\lambda_{\rm 4H}}\lambda$ denote the Higgs self-couplings that could be modified by new physics.
We have added renormalization constants $Z_{\kappa_{\rm 3H} }$ and $Z_{\kappa_{\rm 4H} }$ for the coupling modifiers $\kappa_{\lambda_{\rm 3H}}$ and $\kappa_{\lambda_{\rm 4H}}$ to account for potential new physics effect in renormalization.
Following the general principle of the $\kappa$ framework, we have assumed that the new physics does not affect the vacuum expectation value and Higgs mass when we rescale the Higgs self-couplings.
The second term in the first line of Eq. (\ref{eq:Lsm}) does not contain any field and thus can be safely dropped.
Writing $Z=1+\delta Z$, the third term can be expanded as
\begin{align}
(\mu^2 v - \lambda v^3)H + 
   [ (\delta Z_{\mu^2}+\delta Z_{\phi}+\delta Z_v)\mu^2 v - ( \delta  Z_{\lambda} + 2\delta Z_{\phi}+3\delta Z_v ) \lambda v^3 ]H +\cdots
\end{align}
where we have neglected higher-order corrections that are products of two $\delta Z$'s.
We choose the renormalization condition such that there is no tadpole contribution.
This condition requires $\mu^2  = \lambda v^2$ at the tree level,
and 
$     (\delta Z_{\mu^2}-\delta  Z_{\lambda}-\delta Z_{\phi} -2\delta Z_v )\mu^2 v + T =0$
at the one-loop level
with $T$ being the contribution from the one-loop tadpole diagrams.
The vacuum expectation value appears always in the form of $Z_{\phi}^{1/2}Z_v v $, and is closely related to the massive gauge boson mass. Therefore $\delta Z_v$ would be determined only after considering the renormalization of the EW gauge sector.
Since we focus on the corrections induced by the Higgs self-couplings, we can simply take $\delta Z_v + \delta Z_{\phi}/2 = 0$.
We adopt dimensional regularization, i.e., setting the space-time dimension $d=4-2\epsilon$, to regulate the ultraviolet divergence
and take $\mu_R$ as the renormalization scale.
The tadpole diagram is evaluated to be 
\begin{align}
    T = \frac{3 \lambda_{\rm 3H} v}{16\pi^2}m_H^2\left(\frac{1}{\epsilon}+\ln\frac{\mu_R^2}{m_H^2}+1\right)\,.
    \label{eq:tadpole}
\end{align}

The mass $m_H$ of the Higgs boson can be determined from the quadratic term  in Eq. (\ref{eq:Lsm}), 
\begin{align}
 & \frac{1}{2}(\partial_{\mu}H)^2 -\mu^2 H^2+\frac{1}{2}\delta Z_{\phi} (\partial_{\mu}H)^2 - \left(\frac{3}{2}\delta  Z_{\lambda}+\frac{5}{2}\delta Z_{\phi} -\frac{1}{2}\delta Z_{\mu^2} + 3\delta Z_v \right)\mu^2 H^2 \nonumber\\
 \equiv  & \frac{1}{2}(\partial_{\mu}H)^2 -\frac{1}{2}m_H^2 H^2 
 +\frac{1}{2}\delta Z_{\phi} (\partial_{\mu}H)^2
 - \frac{1}{2}(\delta Z_{m_H^2} + \delta Z_{\phi} ) m_H^2 H^2 \,.
\end{align}
On the right-hand side, we have introduced the classical mass terms.
 By comparing both sides, it is straightforward to get $m_H^2   = 2\mu^2$ and
 $ \delta Z_{m_H^2}  \equiv  \frac{3}{2}\delta  Z_{\lambda}+\frac{3}{2}\delta Z_{\phi} -\frac{1}{2}\delta Z_{\mu^2}  + 3\delta Z_v$.
We choose the on-shell renormalization condition for the Higgs field and obtain
\begin{align}
       \delta Z_{m_H^2} & =\frac{3\lambda_{\rm 4H}}{16\pi^2}\left(\frac{1}{\epsilon}+\ln\frac{\mu^2_R}{m_H^2}+1\right)
       +\frac{9\lambda_{\rm 3H}^2 v^2}{m_H^2}\frac{1}{8\pi^2}\left(\frac{1}{\epsilon}+\ln\frac{\mu^2_R}{m_H^2}+2-\frac{\pi}{\sqrt{3}}\right), \nonumber\\
       \delta Z_{\phi} & =\frac{9\lambda_{\rm 3H}^2 v^2}{8\pi^2}\frac{\sqrt{3}-2\pi/3}{\sqrt{3}m_H^2}\,.
\end{align}

Combining the above equations, we derive the results for the other renormalization constants, $ \delta Z_{\mu^2}$ and $\delta Z_{\lambda}$.
Then the counter-term for the triple Higgs interaction in Eq.~(\ref{eq:Lsm}) is given by 
\begin{align}
    \delta \lambda_{\rm 3H} & \equiv \delta Z_{\lambda}  + 2\delta Z_{\phi}+\delta Z_{v} +\delta Z_{\kappa_{\rm 3H} } \nonumber \\
   &  =-\frac{3\lambda_{\rm 3H}}{16\pi^2}\left(\frac{1}{\epsilon}+\ln\frac{\mu_R^2}{m_H^2}+1\right)+ \frac{3\lambda_{\rm 4H}}{16\pi^2}\left(\frac{1}{\epsilon}+\ln\frac{\mu_R^2}{m_H^2}+1\right) \nonumber \\
  & +\frac{3\lambda_{\rm 3H}^2 v^2}{16\pi^2m_H^2}\left(\frac{6}{\epsilon}+6\ln\frac{\mu_R^2}{m_H^2}+21-4\sqrt{3}\pi\right) 
    +\delta Z_{\kappa_{3H} }
\end{align}
with $\delta Z_{\kappa_{\rm 3H} }\equiv Z_{\kappa_{\rm 3H} }-1$.

Including the contribution of counter-terms, we obtain the result with explicit Higgs self-coupling dependence for the one-particle reducible diagrams,
\begin{align}
&\mathcal{M}_{gg\to H^*\to HH}^{\rm LO}\times  \bigg\{
\frac{3}{16\pi^2}\frac{1}{\epsilon}\left(-2\lambda_{\rm 4H}-\lambda_{\rm 3H}+6\lambda_{\rm 3H}^2\frac{v^2}{m_H^2}\right)  +\delta Z_{\kappa_{\rm 3H} } \nonumber\\
&+\frac{3}{16\pi^2}\ln\frac{\mu_R^2}{m_H^2}\left[-2\lambda_{\rm 4H}-\lambda_{\rm 3H}+6\lambda_{\rm 3H}^2\frac{v^2}{m_H^2}\right] \nonumber \\
&-\frac{ 9\lambda_{\rm 3H}^2 }{8\pi^2}\frac{v^2}{s-m_H^2}\left[\beta\left(\ln\left(\frac{1-\beta}{1+\beta}\right)+i\pi\right)+\frac{s}{m_H^2}\bigg(1-\frac{2\pi}{3\sqrt{3}}\bigg)+\frac{5\pi}{3\sqrt{3}}-1
\right]\nonumber \\
& +\frac{3\lambda_{\rm 3H}^2}{8\pi^2}\frac{v^2}{m_H^2}\left(12-\frac{7\pi}{\sqrt{3}}\right) 
-\frac{9\lambda_{\rm 3H}^2 v^2}{4\pi^2}C_0[m_H^2, m_H^2, s, m_H^2, m_H^2, m_H^2]
\nonumber\\
&- \frac{3\lambda_{\rm 4H}}{16\pi^2}\left[\beta\left(\ln\left(\frac{1-\beta}{1+\beta}\right)+i\pi\right)+5-\frac{2\pi}{\sqrt{3}}\right] -\frac{3\lambda_{\rm 3H}}{16\pi^2}  \bigg\}\,,    
\end{align}
where $s=(p_1+p_2)^2, \beta=\sqrt{1-4m_H^2/s}$ and $C_0[m_H^2, m_H^2, s, m_H^2, m_H^2, m_H^2]$ is a scalar integral that can be calculated by Package-X \cite{Patel:2016fam}. 
$\mathcal{M}_{gg\to H^*\to HH}^{\rm LO}$ is the LO amplitude which contains the Higgs self-coupling.
In the SM, 
the divergences in the first line vanish and $\delta Z_{\kappa_{\rm 3H} }$ is not needed. 
In the general $\kappa$ framework, it is essential to include $\delta Z_{\kappa_{\rm 3H} }$ in renormalization.
The presence of $\delta Z_{\kappa_{\rm 3H} }$ for general Higgs self-couplings demonstrates that the cross section in the $\kappa$ framework can not be derived from the result in the SM, as mentioned in the introduction.
We adopt the $\overline{\rm MS}$ scheme to subtract the divergences.
As a result, the coupling modifier is scale-dependent.
If it is expressed in terms of the value at the scale $m_H$, then an additional contribution from its perturbative expansion cancels the above $\ln (\mu_R^2/m_H^2)$ term exactly.

The $\kappa$ framework was initially established based on the signal strength obtained from experimental results.
Here we have presented a field definition of the $\kappa$ framework in the Higgs sector so that one can perform higher-order calculations.
A more systematical approach is to use the Higgs effective field theory (HEFT) with the electroweak chiral Lagrangian \cite{Buchalla:2013rka}, which can be considered as an upgrade of the $\kappa$ framework to a quantum field theory and provides a general EFT description of the electroweak interactions with the presently known elementary particles under a cutoff scale around a few TeV \cite{Buchalla:2017jlu}.
The physical Higgs field $H$ is introduced as a singlet under $SU(2)_L\times U(1)_Y$ and the chiral symmetry.
Our strategy, which includes the choice of unitary gauge and the implementation of the $\kappa$ parameter in the symmetry-broken phase, 
is equivalent to the application of HEFT in Higgs boson pair production.
This equivalence can be easily verified by comparing our Lagrangian in eq. (\ref{eq:Lsm}) and the one in eq. (2.5) of ref. \cite{Herrero:2022krh}.

\begin{figure}[htbp]
	\centering
       \includegraphics[width=0.48\textwidth]{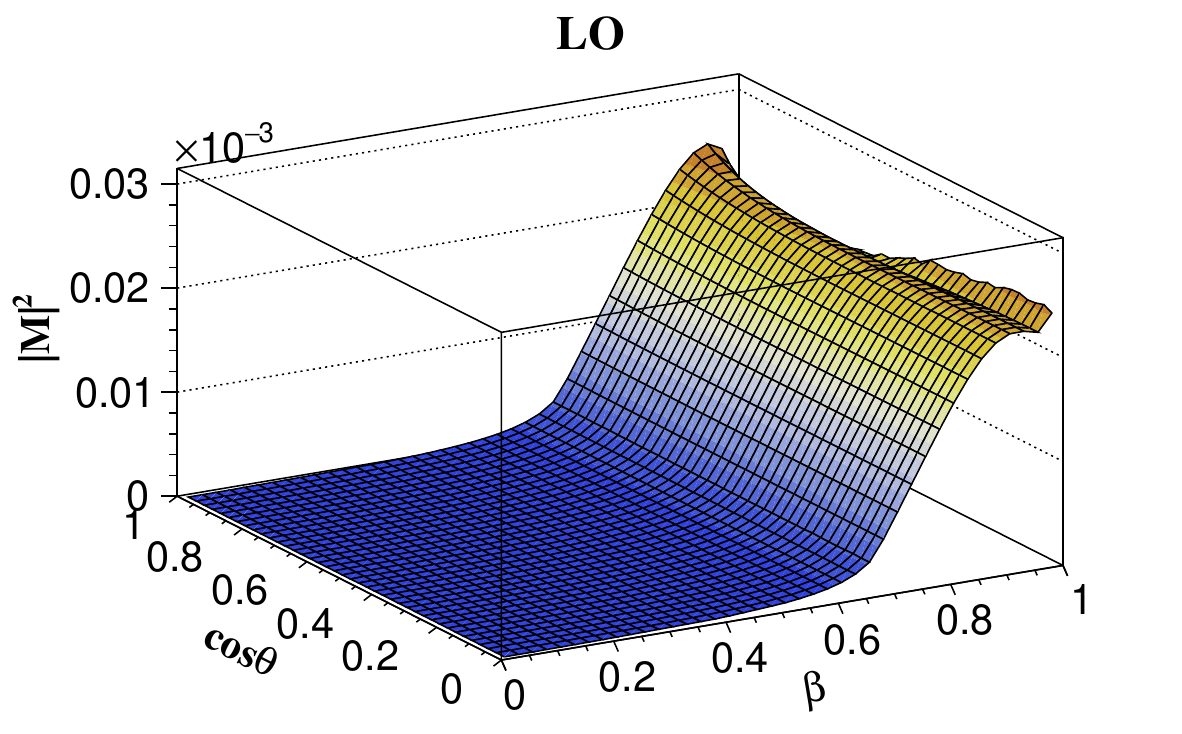}  \includegraphics[width=0.48\textwidth]{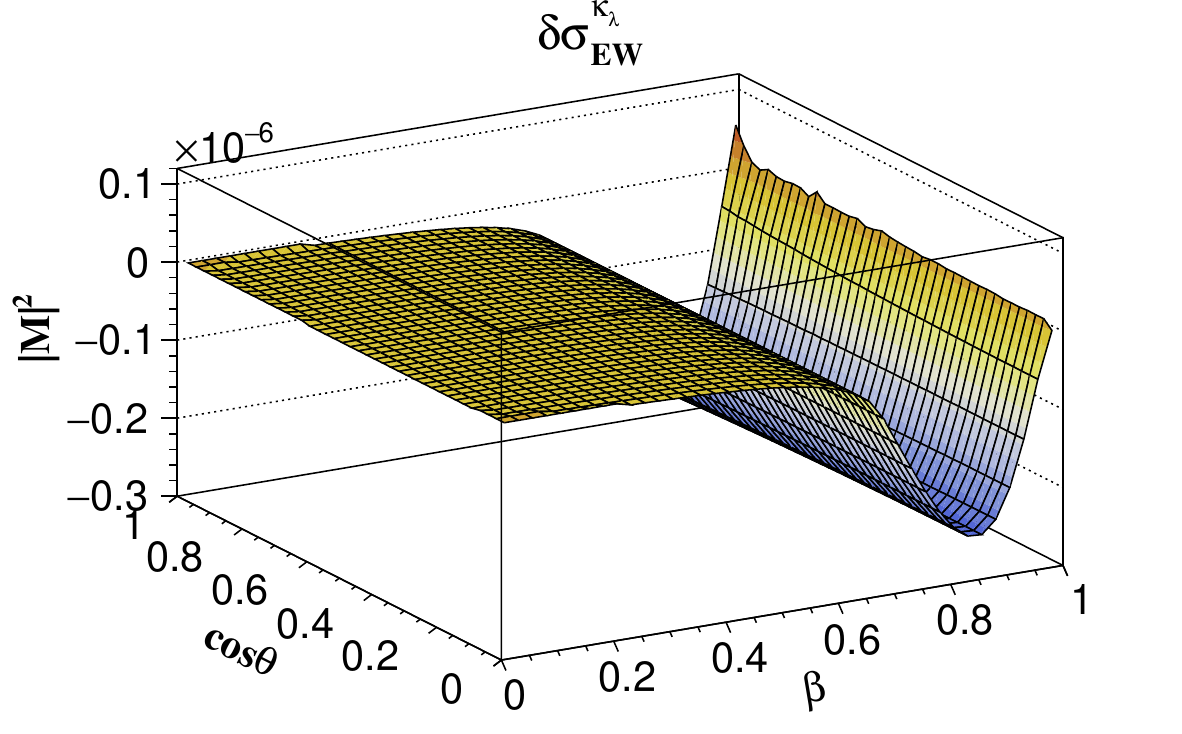}
	\caption{The LO squared amplitudes (left) and $\lambda$ dependent EW corrections (right).}
	\label{figampNLO}
\end{figure}

Now we are ready to compute the finite part of the squared amplitudes.
We set the two-dimensional grid as a function of the Higgs velocity $\beta$ and $\cos\theta$ with $\theta$ the scattering angle.
The range of $\beta$ is from 0 to 1,
and $\cos\theta $ is also taken from 0 to 1 since the squared amplitudes are symmetric under $\theta \to \pi-\theta$.
The grids for the LO squared amplitudes and ${\lambda}$ dependent EW corrections are illustrated in Fig.~\ref{figampNLO}. 
We see that the squared amplitudes are stable against the change within the region $\beta < 0.6$.
For larger $\beta$, the LO squared amplitudes rise dramatically and then start to drop when $\beta > 0.9$.
By contrast, the $\lambda$ dependent correction first decreases and then increases when $\beta$ is larger than 0.85.
These variations are mainly due to the large logarithms $\ln^i (1-\beta)$.
And therefore we have constructed the grid as a function of $\ln^i (1-\beta)$ for $\beta \ge 0.96$.
We have tested the grid by comparing the generated values\footnote{The Lagrange interpolation method is used to generate the squared amplitudes at the phase space point not on the lattice.} and the ones obtained by direct high-precision computation at some phase space points that are not on the grid lattice. 
We find good agreement at the per-mille level.
We have used the grid to calculate the LO total cross section by performing the convolution with the parton distribution function (PDF) and phase space integrations.
Comparing it to the result obtained using analytical expressions or the
\textsc{OpenLoops} package \cite{Cascioli:2011va,Buccioni:2017yxi,Buccioni:2019sur},
we find the relative difference lower than $\mathcal{O}(10^{-3})$.

The LO cross section of the VBF channel also exhibits a quadratic dependence on the trilinear Higgs coupling.
The higher power dependence can be obtained by calculating the one-loop diagrams with an additional Higgs propagator.
The calculation is standard except for the renormalization, which has been elaborated above.
We have implemented our analytical results in the  
 \textsc{proVBFHH} program~\cite{Cacciari:2015jma,Dreyer:2018rfu} and used the \textsc{QCDLoop} package~\cite{Ellis:2007qk} to  evaluate the scalar one-loop integrals.

\section{Numerical results and improved constraints on the Higgs boson self-coupling}  \label{results}

In our numerical calculations, we take $v=(\sqrt{2}\,G_F)^{-1/2}$ with the Fermi constant $G_F=1.16637\times 10^{-5}\,\rm GeV^{-2}$, the Higgs boson mass $m_H=125$ GeV, and the top quark mass $m_t=173$ GeV. 
For the VBF channel, we set the EW gauge boson masses $M_W=80.379$ GeV and $M_Z=91.1876$ GeV.
We use the $\rm PDF4LHC15\_nlo\_100\_pdfas$ PDF set \cite{Butterworth:2015oua}, and the associating strong coupling $\alpha_s$. 
The default renormalization scale in $\alpha_s$ and the factorization scale in the PDF are chosen to be $\mu_{R, F}=m_{HH}/2$ in the ggF channel and  $\mu_{R, F}=\sqrt{-q_i^2}$   in the VBF channel with $m_{HH}$ being the Higgs pair invariant mass and $q_i$ being the transferred momenta from quark lines.

The EW corrections that contain higher power dependence on the Higgs self-coupling are given by
\begin{align} \label{eqkaexp}
\delta\sigma^{\kappa_{\lambda}}_{\rm ggF,EW}  & = (0.067\kappa_{\lambda_{\rm 3H}}^4-0.105\kappa_{\lambda_{\rm 3H}}^3-0.006\kappa_{\lambda_{\rm 3H}}^2\kappa_{\lambda_{\rm 4H}}-0.166\kappa_{\lambda_{\rm 3H}}^2\nonumber\\
&~~~+0.070\kappa_{\lambda_{\rm 3H}}\kappa_{\lambda_{\rm 4H}} -0.149\kappa_{\lambda_{\rm 4H}})~~ {\rm fb}  
\end{align}
for the ggF channel and 
\begin{align}
        \delta\sigma_{\rm VBF,EW}^{\kappa_\lambda}&=(0.0196\kappa_{\lambda_{\rm 3H}}^4-0.0235\kappa_{\lambda_{\rm 3H}}^3-0.0014\kappa_{\lambda_{\rm 3H}}^2\kappa_{\lambda_{\rm 4H}}-0.0165\kappa_{\lambda_{\rm 3H}}^2 \nonumber\\
    &~~~+0.0126\kappa_{\lambda_{\rm 3H}}\kappa_{\lambda_{\rm 4H}}-0.0193\kappa_{\lambda_{\rm 4H}})~~ {\rm fb}
\end{align}
for the VBF channel.
We have computed all the $\mathcal{O}(\lambda_{\rm 3H}^i), i\ge 2$ contributions in the amplitude.
The above $\kappa_{\lambda_{\rm 3H}}^2$ terms arise because 
we want to keep the cancellation relation between the $\mathcal{O}(\lambda_{\rm 3H})$ and $\mathcal{O}(1)$ amplitudes at LO.
The cubic $\kappa_{\lambda_{\rm 3H}}^3$ and quartic $\kappa_{\lambda_{\rm 3H}}^4$ terms appear for the first time up to this perturbative order. 
Though their coefficients are rather small,
they provide notable corrections to the cross section if $\kappa_{\lambda_{\rm 3H}}$ is chosen much larger than 1.
As seen from Table \ref{tab_kappa}, 
the $\lambda$ dependent  corrections in the ggF (VBF) channel reach $88\%$ ($78\%$) of the LO cross section for $\kappa_{\lambda_{\rm 3H}}=6$.

\begin{table}[h]
	\renewcommand\arraystretch{1.5} 
	\centering
	\begin{tabular}{|c|c|c|c|c|c|c|c|}
		\hline
		\multirow{2}{*}{$\kappa_{\lambda_{\rm 3H}}$} & 
		\multirow{2}{*}{$\kappa_{\lambda_{\rm 4H}}$} & \multicolumn{3}{|c|}{ggF} & \multicolumn{3}{|c|}{VBF}\\
		\cline{3-8}
		& & $\sigma_{\rm LO}^{\kappa_{\lambda}}$ &
		$\sigma_{\rm NNLO-FT}^{\kappa_{\lambda}}$ &
		$\delta\sigma^{\kappa_\lambda}_{\rm EW}$ & $\sigma_{\rm LO}^{\kappa_{\lambda}}$ & $\sigma_{\rm NNNLO}^{\kappa_\lambda}$ & $\delta\sigma^{\kappa_\lambda}_{\rm EW}$ \\
		\hline
		1 & 1 & 16.7 & 31.2  & -0.287 &  1.71 & 1.69 & $ -2.85 \times 10^{-2}$ \\
		\hline
		3 & 1 & 8.59 & 18.4 & 1.15 & 
		3.60 & 3.53 &$8.09\times 10^{-1}$ \\
		\hline
		6 & 1 & 67.3 & 161 & 58.7  &   
		25.1 & 24.6 & 19.7\\
		\hline
		1 & 3 & 16.7  & 31.2  & -0.455 & 1.71 & 1.69 & $-4.47 \times 10^{-2}$ \\
		\hline
		1 & 6 & 16.7 & 31.2 & -0.708 & 1.71 & 1.69  &$-6.91 \times 10^{-2}$  \\
		\hline
		3 & 3 & 8.59 & 18.4 & 1.17 & 3.60 & 3.53 & $8.21\times 10^{-1}$ \\
		\hline
		6 & 6 & 67.3 & 161 & 59.1 & 25.1 &  24.6 & 19.7  \\
		\hline
	\end{tabular}
	\caption{Cross sections (in fb) of ggF and VBF Higgs boson pair production for different values of $\kappa_{\lambda_{\rm 3H}}$ and $\kappa_{\lambda_{\rm 4H}}$ at the 
		13 TeV LHC. }
\label{tab_kappa}
\end{table}

In addition, there is a new dependence on the quartic Higgs self-coupling $\lambda_{\rm 4H}$.
Because this dependence is only linear and the corresponding coefficients are small,
their contributions are negligible.
From Table \ref{tab_kappa}, it can be observed that the cross section varies by $0.7\%$ when $\kappa_{\lambda_{\rm 4H}}$ changes from 1 to 6 while keeping $\kappa_{\lambda_{\rm 3H}}=6$. 
As a consequence, we do not expect that a meaningful constraint on the quartic Higgs self-coupling can be extracted from Higgs boson pair production at the LHC.

We can make a comparison with the results in Refs. \cite{Bizon:2018syu,Borowka:2018pxx}.
The authors of these papers have obtained similar expressions for the cross sections in the ggF channel.
However, they have assumed that the triple and quartic Higgs self-couplings are modified by one dimension-six and one dimension-eight operators 
\footnote{There are new physics scenarios, e.g, the Nambu-Goldstone Higgs potential \cite{Kaplan:1983fs}, in which the modification of the Higgs potential can not be described by higher-dimensional operators but can be accommodated in the $\kappa$ framework \cite{Agrawal:2019bpm}. }.
They performed calculations, especially renormalization, in terms of the coefficients of higher-dimensional operators, and then transformed the results onto the basis of $\kappa_{\lambda_{\rm 3H}} $ and $\kappa_{\lambda_{\rm 4H}}$.
These results can not be directly compared with the experimental analysis in the $\kappa$ framework.

\begin{figure}
    \centering
    \includegraphics[width=8cm]{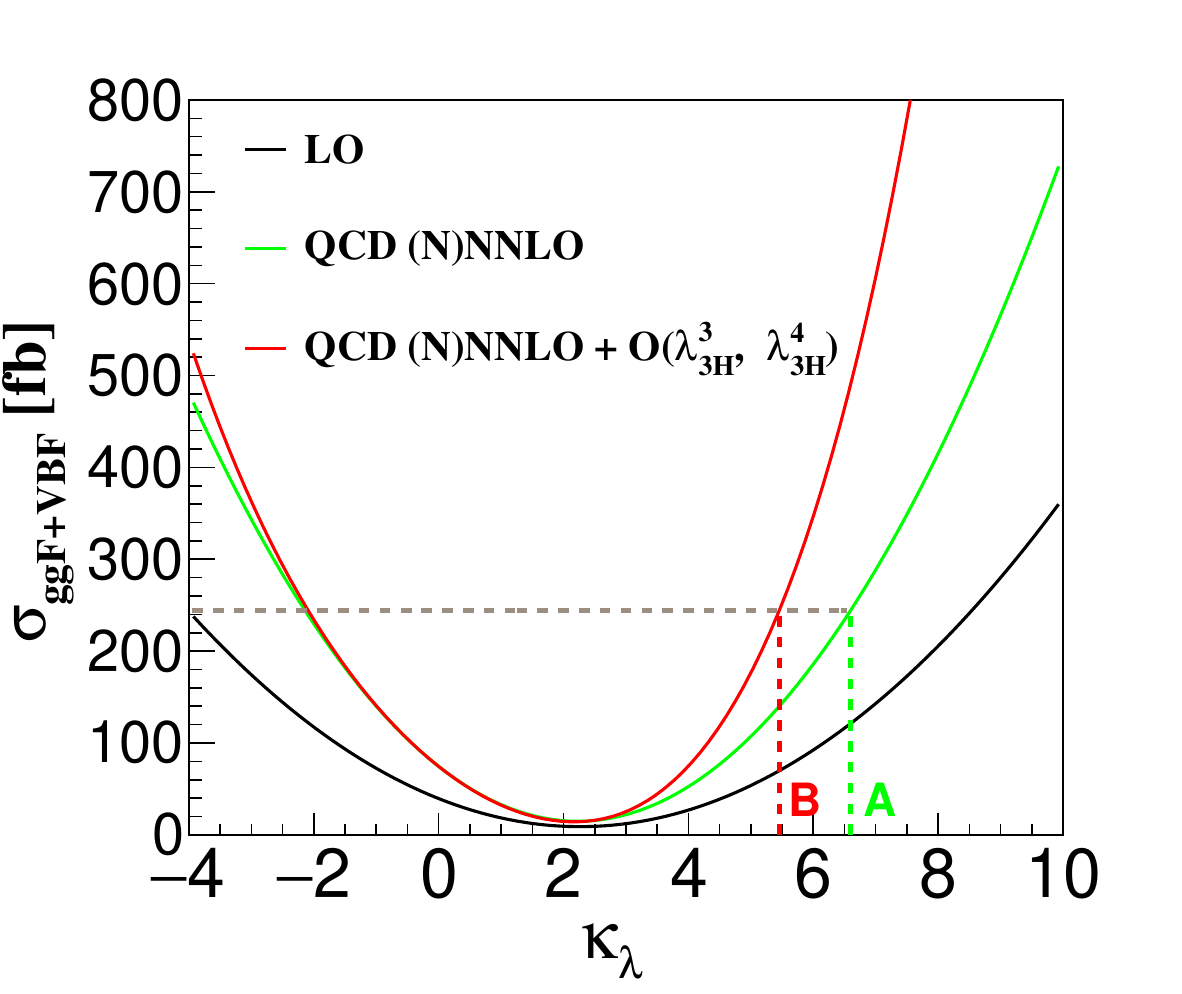}
    \caption{Cross sections of Higgs boson pair production  at the 13 TeV LHC including both ggF and VBF processes as a function of $\kappa_{\lambda_{\rm 3H}}=\kappa_{\lambda_{\rm 4H}}=\kappa_{\lambda}$.  The black line represents the LO result, and the green line denotes the result with (N)NNLO QCD corrections in the ggF (VBF) channel.  
    The red line indicates the result including higher power dependence on the Higgs boson self-coupling.  
    The current and improved upper limits are labeled by the points A and B, respectively. }
    \label{fig:kappa}
\end{figure}

In Fig. \ref{fig:kappa}, we show different perturbative predictions for the cross sections of both ggF and VBF Higgs boson pair production at the 13 TeV LHC as a function of $\kappa_{\lambda_{\rm 3H}}=\kappa_{\lambda_{\rm 4H}}=\kappa_{\lambda}$.
It is evident that higher-order perturbative corrections dramatically change the functional form.
The current experimental upper limit by the ATLAS (CMS) collaboration on $\kappa_{\lambda}$ is $6.6$ (6.49) based on the theoretical predictions at QCD NNLO in the ggF channel and NNNLO in the VBF channel; see Table \ref{tab_kappa}.
Taking $\delta\sigma^{\kappa_\lambda}_{\rm EW}$ corrections into account and
assuming that the QCD and EW corrections are factorizable, the upper limit is narrowed down to 5.5 (5.39).
These limits are almost the same when keeping $\kappa_{\lambda_{\rm 4H}}=1$.
If the scale uncertainties are considered \cite{Baglio:2020wgt}, the upper limit spans in the range $(6.5,6.8)$ in the ATLAS result, which would decrease to $(5.4,5.6)$ after including higher power dependence. 
In the CMS result, the upper limit changes from (6.40, 6.67) to (5.33, 5.50).
The lower limits are only modified slightly.

\begin{figure}[htbp]
	\centering
	\includegraphics[width=0.32\textwidth]{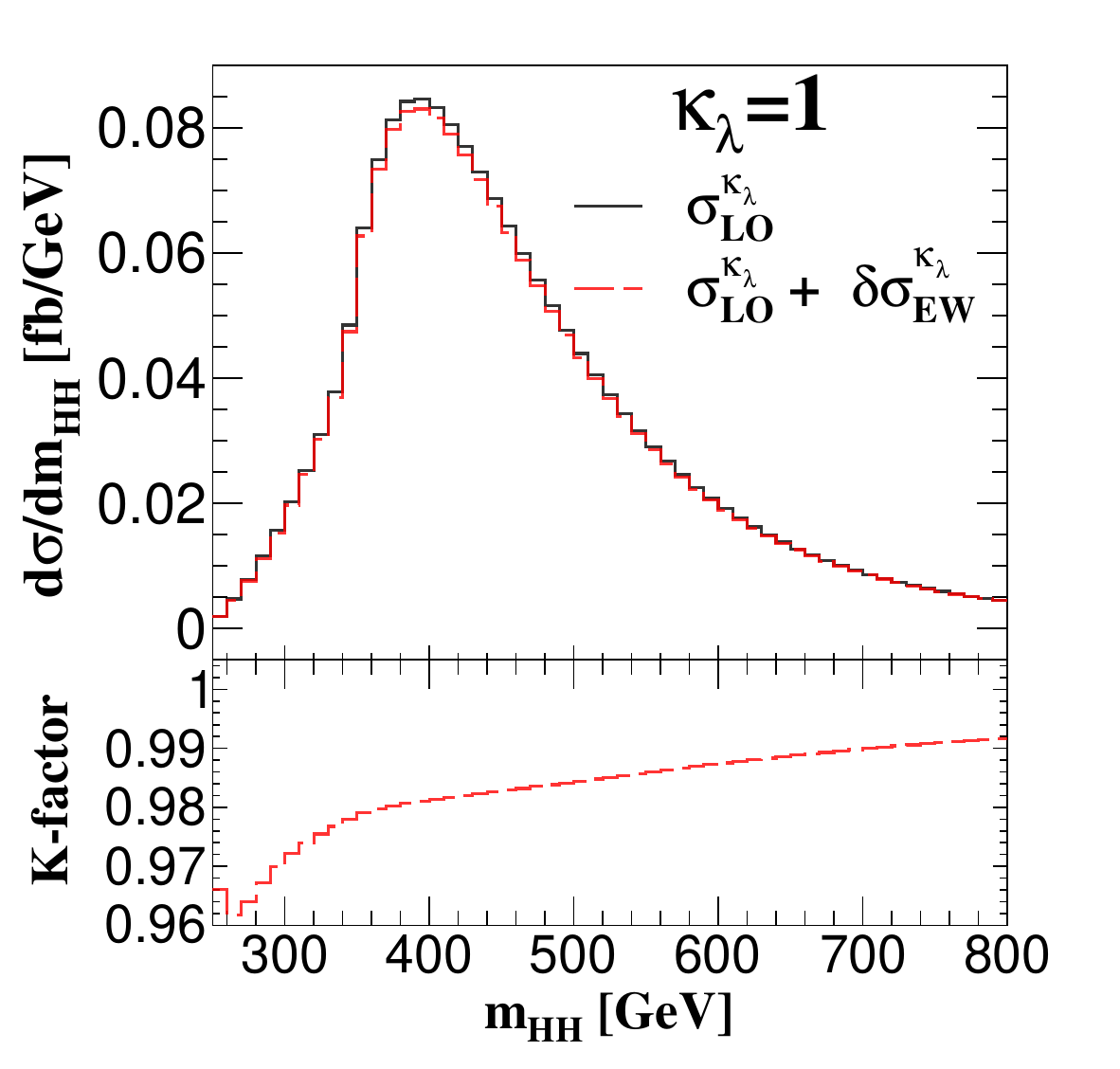}
	\includegraphics[width=0.32\textwidth]{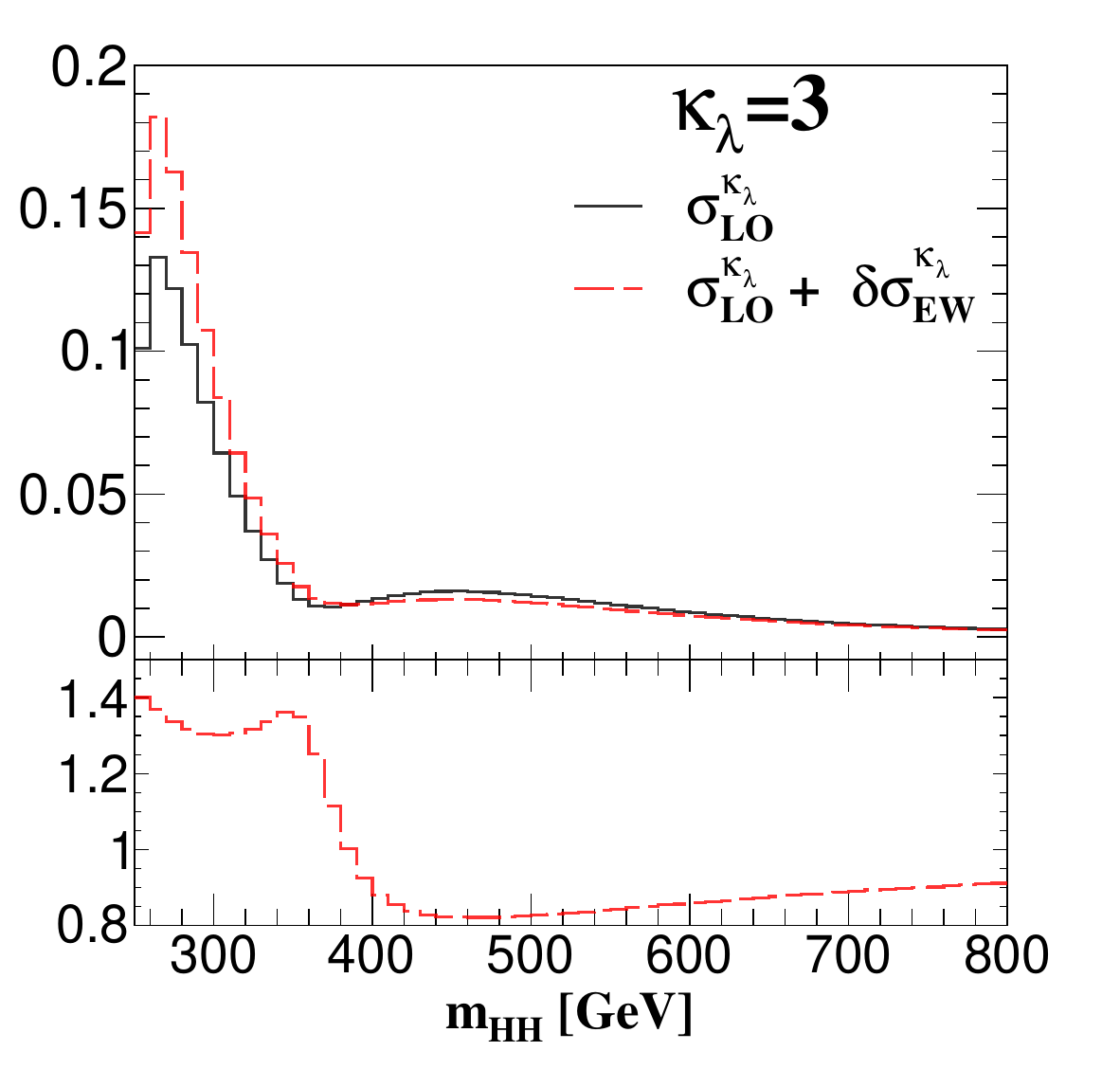}
	\includegraphics[width=0.32\textwidth]{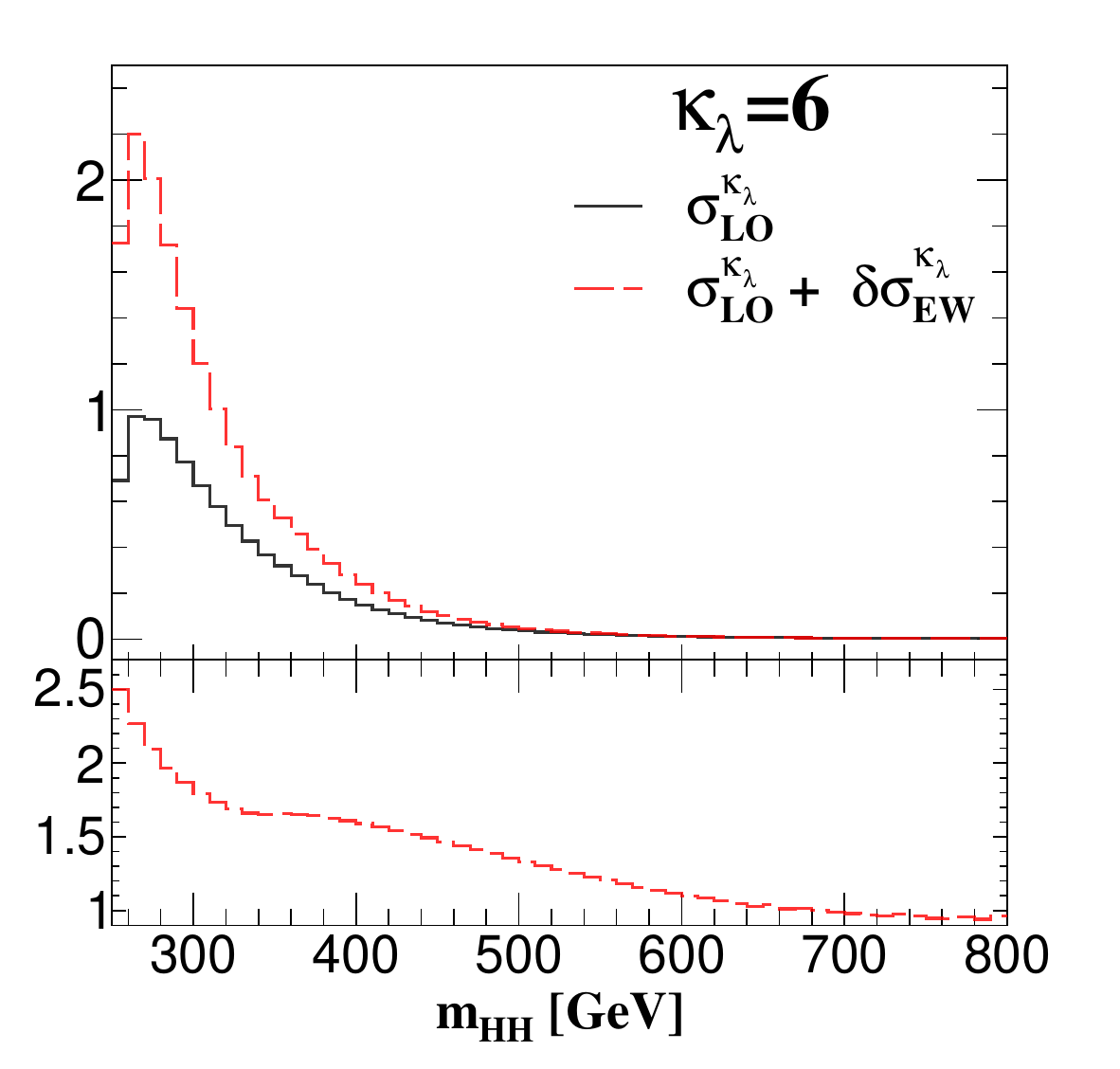}
\caption{The Higgs boson pair invariant mass distributions at LO and with $\delta\sigma_{\rm EW}^{\kappa_\lambda}$ corrections at the 13 TeV LHC. We have used  $\kappa_{\lambda_{\rm 3H}}=\kappa_{\lambda_{\rm 4H}}=\kappa_{\lambda}$. }
\label{sig_kappa_3}
\end{figure}

Lastly, the Higgs boson pair invariant mass $m_{HH}$ distributions are shown in Fig.~\ref{sig_kappa_3}. 
The peak position moves from 400 GeV to 260 GeV when $\kappa_{\lambda}$ varies from 1 to 6,
which indicates the Higgs bosons tend to be produced 
with very small velocity in the case of large $\kappa_{\lambda}$ values.
This feature could help to set optimal cuts in the experimental analysis to enhance the sensitivity.
From this figure, we also observe that the $\lambda$ dependent corrections have a great impact on the distributions, especially in the small $m_{HH}$ region, and therefore they should be included in future studies.

\section{Conclusion} \label{conclusion}

The precise shape of the Higgs potential stands as a fundamental enigma in particle physics.
The current limits on the Higgs self-coupling are predominantly derived based on the assumption that the cross section of the  Higgs boson pair production is a quadratic function of the self-coupling.
We find that the function form should be generalized to include quartic and cubic power dependencies on the Higgs self-coupling 
which arise due to higher-order quantum corrections induced by virtual Higgs bosons.

We propose a proper renormalization procedure to explicitly retain the Higgs self-couplings at each calculation step and introduce renormalization of the coupling modifiers to ensure the cancellation of all ultraviolet divergences.
We present numerical results of the cross sections of both the ggF and VBF channels at the LHC including higher power dependencies on the Higgs self-coupling.
With these improved function forms, we demonstrate that the upper limit set by the ATLAS (CMS) collaboration on the trilinear Higgs self-coupling normalized to its SM value
is reduced from 6.6 (6.49) to 5.5 (5.39).
This more precise constraint is achieved without more data being analyzed, underscoring the critical importance of  incorporating  higher power dependencies on the Higgs self-coupling in the cross section.

Furthermore, we find it hard to derive any useful constraint on the quartic Higgs self-coupling solely from Higgs boson pair production. 
To probe the quartic self-coupling, alternative channels such as triple Higgs boson production may be explored, necessitating collider facilities with higher energies than the LHC to unveil insights into this aspect of the Higgs potential.


\section*{Acknowledgments}
We express our gratitude to Huan-Yu Bi, Yan-Qing Ma, and Huai-Min Yu for comparing the numerical results of the two-loop amplitudes.
We also thank Shan Jin, Yefan Wang and Lei Zhang for helpful discussion.
This work was supported in part by the National Natural Science Foundation of China under grant No. 12275156, No. 12321005, No. 12375076 and the Taishan Scholar Foundation of Shandong province (tsqn201909011).

\end{sloppypar}

\bibliographystyle{JHEP}
\bibliography{ref}

\end{document}